\documentstyle[12pt,a4]{article}

\makeatletter
 
 \@addtoreset{equation}{section}
\makeatother

\begin{document}


\title{Multipole Expansion Model in Gravitational Lensing}
\author{Takeshi FUKUYAMA \footnotemark[1],
        Yuuko KAKIGI \footnotemark[1],
         and Takashi OKAMURA \footnotemark[2] \\
        $\ast$~~Department of Physics, \\
        Ritsumeikan University, Kusatsu, \\
        Shiga, 525 Japan \\
        $\dagger$~~Department of Physics, \\
        Tokyo Institute of Technology, \\
        Oh-Okayama, Meguro-ku, Tokyo, 152 Japan}
\date{}

\maketitle


\begin{abstract}

Non-transparent models of multipole expansion model and two point-mass
model are analyzed from the catastrophe theory.
Singularity behaviours of $2^n$-pole moments are discussed.
We apply these models to triple quasar PG1115+080 and compare with
the typical transparent model, softened power law spheroids.
Multipole expansion model gives the best fit among them.


\end{abstract}


\vfill\eject


\section{Introduction}

Gravitational lensing is and will be playing very important roles in
astrophysics \cite{Schneider} \cite{Blandford}.
It covers the large region in scale from the brown dwarf to the supercluster
of galaxies.
Microlensing uncovers the dark matter in the halo and the planetary
system in the neighbourhood of the Sun \cite{Roulet}.
In this letter we deal with a special  macrolensing, that is, the multiple
quasar supposed to be caused by a single galaxy in the foreground.
It enables us to reveal the structure of the lensing galaxy without 
the ambiguity of surrounding galaxies.
It also gives us the information of the large scale structure of the
universe through the angular diameter distance.
\par
One of the most popular models of lensing galaxy is the spheroid
with softened power law behaviours \cite{Binney}.
The physical implication of this model is very clear and it is
easy to compare with observations, such as luminosity distribution
and ellipticity, etc.
However these merits are also the demerits of this model
since it is too restrictive to model shape.
We are not sure whether the lensing galaxy is elliptical or spiral.
It may be even complex system composed of several galaxies.
This model is also analytically non-tractable.
On the other hand multipole expansion model
(hereafter MPE model)
is more general than the spheroid model.
MPE model is applicable to any complex system
as far as the lensing objects are compact with respect to impact parameters.
Also it is analytically tractable.
\par
Using it, we are able to estimate the structure of the lensing object 
systematically.
That is, the more we incorporate multipole components,
the more we can know the detailed structure of the lensing object.
\par
From the observational side, gravitational lensing effect seems to be
very promissing.
Deep space survey by HST is now working and Sloan Digitized Sky Survey
will work soon \cite{SDSS}.
However the present status of observation is not still sufficient
to uncover the affirmative results.
For instance,
parameter fitting is very sensitive to the image positions
\cite{Katsukun}
whose observation cannot be precise in many cases.
We need more detailed and more precise data 
to be adjusted.
In such a  situation
it is also very useful to study the same lensing phenomena by several
models, which reveal the common and therefore true features of
lensing objects.
\par
In the previous paper \cite{prev} we discussed about the relationship
between multipole moments and caustic singularities,
and applied it to PG1115+080.
In this treatise it is also shown that complex coordinates are very useful.
\par
The merit of MPE model is that we can probe into the detailed
structure systematically by steping up higher order of multipoles.
An arbitrarily complex but compact lensing object is analytically tractable
as far as each $2^n$-pole moment is discussed separately.
So in this article we study the structure of $2^n$-pole term systematically.
If we sum up all contributions, its treatment goes beyond analytical survey
and is devoted to the numerical calculations.
\par
Two point-mass lens is situated at the antipodal point of multipole expansion
model.
\par
Two point-mass model consists of a series of multipole moments.
However it allows analytical survey as a whole
though a straightforward application of catastrophe theory
still needs numerical calculations.
In order to understand the essential physical implication of any event,
analytical survey is indispensable.
\par
We apply the above mentioned models to PG1115+080 and discuss their common
and discrepant characters among them.
\par
This paper is organized as follows.
In section two we review the metamorphoses of caustic for later use
whose detailed arguments are given in our previous work \cite{prev}.
This metamorphoses are applied to MPE model in section three
and to the two point-mass model in section five.
Section four is devoted to the analysis of the detailed structure of
$2^n$-pole contribution in MPE model.
MPE model is applied to the multiple lens system of PG1115+080
in section six.
The obtained numerical results are compared with other models,
especially with elliptical lens model in section seven.
Section eight is devoted to discussions.
 

\section{Metamorphoses of the Caustics}

For later studies, we summarize the metamorphoses of the caustics.
This is mainly based on our previous work \cite{prev}.
\par
Using complex variables, singularities are classified in compact form.
\par
Critical line is

\begin{eqnarray}
D &\equiv& -4  \hbox{det} (\phi_{\alpha \beta}) \nonumber \\
  &=& 4(|\phi_{z \bar z}|^2 - |\phi_{z z}|^2) = 0~~,
 ~~~~~~(\alpha, \beta = z, \bar z)
\label{gecrit}
\end{eqnarray}

where $\phi$ is Fermat potential and $z$ is complex representation of image
position, $z = x_1 + i x_2$.
$\bar z$ is complex conjugate of $z$.
Subscripts $\alpha$, $\beta$ on $\phi$ denotes the derivative 
$\partial_\alpha \partial_\beta \phi$ and so on.
\par
Here we use the Poisson equation

$$ \phi_{z \bar z} = {1 \over 2} (1-\kappa)~~. $$

$\kappa$ is the normalized surface mass density
and is equal to zero on the light path for non-transparent model
which we will consider in this article.
\par
Then the critical line is reduced to

\begin{equation}
\phi_{z z} \pm  {1 \over 2} = 0
\label{cocrit}
\end{equation}

by the diagonalization in real coordinates.
$+$ ($-$) corresponds to $\phi_{11}=0$ ($\phi_{22}=0$).
\par
Cuspoid sequences are characterized by

\begin{equation}
\hat L^n \phi_z = 0 ~~~~~(n = 1, 2, 3)~~.
\end{equation}

$\hat L$ is the derivative operator along the critical line, 
which is given by

\begin{equation}
\hat L = {i \over 8} (\partial_{\bar z} D \partial_z
                      - \partial_z D \partial_{\bar z})~~.
\end{equation}

Their explicit forms are

\begin{equation}
{\partial^3 \phi \over \partial z^3} \pm
{\partial^3 \phi \over \partial \bar z^3} = 0
\label{cocusp}
\end{equation}

for the cusp ($n = 1$) singularity,

\begin{equation}
({\partial^4 \phi \over \partial z^4} +
  {\partial^4 \phi \over \partial \bar z^4}) \pm
6 ({\partial^3 \phi \over \partial z^3})^2 = 0
\label{coswallow}
\end{equation}

for the swallowtail ($n = 2$) and

\begin{equation}
({\partial^5 \phi \over \partial z^5} \pm
  {\partial^5 \phi \over \partial \bar z^5}) \pm
20 {\partial^3 \phi \over \partial z^3}
\{\ {\partial^4 \phi \over \partial z^4} \pm
     3 ({\partial^3 \phi \over \partial z^3})^2 \}\ = 0
\label{cobutterfly}
\end{equation}

for the butterfly ($n = 3$).
\par
It should be noticed that
the higher cuspoids must also satisfy the conditions of the lower ones.
For instance, butterfly must satisfy Eqs. (\ref{cocrit}), (\ref{cocusp})
and (\ref{coswallow}) as well as Eq. (\ref{cobutterfly}). 
\par
Umbilic is

\begin{equation}
\phi_{\alpha \beta} = 0~~.
\end{equation}

As is easily checked this singularity is not realized for 
the non-transparent model.
\par
Beak-to-beak and lips are characterized by

\begin{equation}
\partial_z D = 0~~,
\label{beaklip}
\end{equation}

\begin{equation}
\hbox{a) Lips} ~~~ \Delta \equiv -\hbox{det} D_{\alpha \beta} > 0
\end{equation}

\begin{equation}
~~\hbox{b) Beak-to-beak}~~~~ \Delta < 0~~.
\end{equation}

The explicit form of Eq.(\ref{beaklip}) is

\begin{equation}
{\partial^3 \phi \over \partial z^3} = 0~~.
\label{lipbeaks}
\end{equation}

$\Delta$ takes the form

\begin{equation}
\Delta =- 4|{\partial^4 \phi \over \partial z^4}| < 0
\end{equation}

for non-transparent model.
Therefore Eq.(\ref{lipbeaks}) leads to beak-to-beak.




\def\half{{1 \over 2}}
\def\Qn{Q_n}
\def\an{\alpha_n}
\def\bn{\beta_n}
\def\expon{1+{2 \over n}}
\def\Im{\hbox{Im}}
\def\Re{\hbox{Re}}
\section{Multipole Expansion (MPE) Model}
Fermat's potential is
 
\begin{equation}
	\tilde\phi=\half |z-z_s|^2 -{1 \over 2 \pi}\int d^2x'~ \kappa(x')~
	           \ln |z-z'|^2~.
\end{equation}

Using the expansion form of logarithmic function,

\begin{eqnarray}
	\ln |z-z'|^2 &=& \ln |z|^2+\ln |1-{z' \over z}|^2
\nonumber \\
	&=& \ln |z|^2-\sum_{n=1}{1 \over n}
	\Bigl\{ \Bigl({z' \over z} \Bigr)^n + c.c. \Bigr\}~, 
\end{eqnarray}

we obtain the multipole expansion form of Fermat's potential,

\begin{eqnarray}
\label{fermul}
	\tilde\phi &=& \half |z-z_s|^2 -{m \over 2} \ln |z|^2
	+\half \sum_{n=1} \Bigl( {q_n \over z^n} + c.c. \Bigr)~,   \\
\label{massmul}
	m &\equiv& {1 \over \pi} \int d^2x~\kappa(x)~,   \\
\label{polemul}
	q_n &\equiv& {1 \over \pi n} \int d^2x~\kappa(x)~z^n~. 
\end{eqnarray}

Thus MPE model becomes a non-transparent model from the convergence
requirement of Taylor expansion.
This property seems to have interrupted popularity in gravitational
lensing. However this model is very useful and applicable to many
lensing events and deserves further investigations 
as will be discussed.

We can transform the potential into,

\begin{eqnarray}
\label{potential}
	&\phi& \equiv {\tilde\phi \over m} =
	\half |u-v|^2 -\half \ln |u|^2
	+\half \sum_{n=1} \Bigl( {Q_n \over u^n} + c.c. \Bigr)~,  
\\
\label{renormalize}
	&u& \equiv {z \over m^\half}~,\qquad 
	v \equiv {z_s \over m^\half}~,\qquad 
	Q_n \equiv {q_n \over m^{1+{n \over 2}}}~. 
\end{eqnarray}

For this potential, critical condition and cusp condition become

\begin{equation}
	|u|^2 =|1+\sum \bn{Q_n \over u^n}|~,
\label{crit}
\end{equation}
\begin{equation}
  \Im \Bigl[\Bigl( 1+\sum \an\bn
	{\bar Q_n \over \bar u^n} \Bigr)^2
	\Bigl( 1+\sum \bn{Q_n \over u^n} \Bigr)^3 \Bigr] =0~,
\label{cuspI}
\end{equation}
\begin{equation}
  \Re \Bigl[\Bigl( 1+\sum \an\bn
	{\bar Q_n \over \bar u^n} \Bigr)^2
	\Bigl( 1+\sum \bn{Q_n \over u^n} \Bigr)^3 \Bigr] >0~,
\label{cuspR}
\end{equation}

where $\an \equiv (n+2)/2$~ and~ $\bn \equiv n(n+1)~$.
Im (Re) denotes the imaginary (real) part.


\section{The Properties of $2^n$- Pole}


Unfortunately, if we consider the terms up to quadrupole moment,
MPE model is already beyond analytical calculation.
However, analytical survey is desirable 
for the essential physical interpretation.
So we consider the case of $2^n$-pole only, 
which allows analytical survey.

Introducing the variables,
$$
	{u^n \over \Qn}=\sqrt{X} e^{i n \theta}~,\qquad
	Y=\sqrt{X}\cos n\theta~,
$$
Eqs.(\ref{crit}), (\ref{cuspI}) and (\ref{cuspR}) are rewritten into,

\begin{eqnarray}
\label{critII}
	&|\Qn|^{4 \over n}X^{\expon}=X+2\bn Y+\bn^2~,
& \\
	&\Im (\hbox{cusp}) \equiv \sin n\theta 
	[(2\an-3)X^2+2\bn (\an^2-3)XY
& \nonumber \\
\label{cuspIm}
	& \qquad +\bn^2 (3\an^2-6\an+1)X-\bn^2 (2Y+\an\bn)^2]=0~,
& \\
	&\Re (\hbox{cusp}) \equiv 
	X^3+(2\an+3)\bn X^2Y+\bn^2\{2(\an^2+3)XY^2-(\an^2-6\an+3)X^2\}
& \nonumber \\
\label{cuspRe}
	&+3\bn^3(\an^2+2\an-1)XY+\an\bn^4(3\an-2)X+\bn^3Y(2Y+\an\bn)^2>0~,
& \\
\label{condXY}
	&X \ge Y^2~.
&
\end{eqnarray}

The last condition follows from the just definition of $X$ and $Y$.
As for the solution of $(X,Y)$, there are $2n$ solutions for $X > Y^2$ 
on lens plane and $n$ solutions for $X=Y^2$.

At first, we examine the cusp conditions, Eqs.(\ref{cuspIm})
and (\ref{cuspRe}), and Eq.(\ref{condXY}).

When $\sin n\theta=0$, $i.e.$, $X=Y^2$, Eq.(\ref{cuspRe}) becomes
$$
	\Re (\hbox{cusp})=Y(Y+\an\bn)^2 (Y+\bn)^3>0~.
$$
Therefore a part of region that satisfies the cusp conditions is
$X=Y^2$ for $Y>0$ or $Y<-\bn$ ($Y \ne -\an\bn$).
Further at the points, $(X, Y)=(\bn^2, -\bn)~, 
(\an^2\bn^2, -\an\bn)$, the beak-to-beak condition is held.

When $\sin n\theta \ne 0$, Eq.(\ref{cuspIm}) becomes

\begin{equation}
	(2\an-3)X^2+2\bn (\an^2-3)XY
	+\bn^2 (3\an^2-6\an+1)X-\bn^2 (2Y+\an\bn)^2=0~.
\label{cuspImII}
\end{equation}

After some calculations, we can obtain the region that satisfies 
Eq.(\ref{cuspRe}) and Eq.(\ref{condXY}),

\begin{equation}
	X>0~, \qquad {\an\bn \over 2\an-3}\ge Y >-\bn~.
\label{region}
\end{equation}

We call the characteristic points, $P_1$, $P_2$ and $P_3$ defined by

$$
	P_1=(\bn^2,~-\bn)~, \qquad P_2=(\an^2\bn^2,~-\an\bn)~,
$$  $$
	P_3=\Bigl({\an^2\bn^2 \over (2\an-3)^2}~,
		~{\an\bn \over 2\an-3}\Bigr)~.
$$

The region in which the cusp conditions is satisfied is depicted 
in Fig.1 for $n=2$. 

\begin{center}
  ------------  \\
     Fig.1      \\
  ------------
\end{center}

For the other values of $n$, topological 
structure does not change.
\par
Next we examine the qualitative behaviour of the critical condition, 
Eq.(\ref{critII}).

Firstly, irrespective of the value of $|\Qn|$, the curve Eq.(\ref{critII}) 
passes through the definite point,
$$
	(0~,~-{\bn \over 2})~,
$$
and passes above $P_1$,
$$
	Y(X=\bn^2)=-\bn+|\Qn|^{4/n}{\bn^{1+4/n} \over 2}> -\bn~.
$$
For $X>0$, the curve has a locally minimum point and 
$Y$ tends to $\infty$ at $X=\infty$.
Furthermore, for fixed $X$, the value of $Y$ increases as
$|\Qn|$ does.

From the above arguments, the number of the solutions of 
the critical condition 
and the cusp conditions changes when the curve Eq.(\ref{critII})
\begin{itemize}
\item[case~(0)] is tangential to $X=Y^2$ 
for $Y<-\bn$ or,
\item[case~(1)] passes through $P_3$~, or
\item[case~(2)] is tangential to $\Im (\hbox{cusp})=0$ for Eq.(\ref{region}).
\end{itemize}
Each case is depicted in Fig.2 for $n=2, 4$.

\begin{center}
  --------------  \\
     Fig.2(a)    \\
  --------------
\end{center}
\begin{center}
  --------------  \\
     Fig.2(b)    \\
  --------------
\end{center}

For the case(0), the value of $|\Qn|$ is 

\begin{equation}
	|\Qn|=|\Qn^{(0)}|\equiv {2 \over n(n+1)(n+2)}
	\Bigl({n \over n+2} \Bigr)^{{n \over 2}}~,
\label{eq}
\end{equation}

and the tangential point is $P_2$.
Therefore beak-to-beak singularity appears.

For the case(1), the value of $|Q_n|$ is

\begin{equation}
	|\Qn|=|\Qn^{(1)}|\equiv {2\an-3 \over \an\bn}
	\Bigl(3{\an-1 \over \an} \Bigr)^{{n \over 2}}
	=3^{n/2}(n-1) |\Qn^{(0)}|~.
\label{qI}
\end{equation}

As known from graphical consideration, then cusps split into
more cusps or merge to less cusps.
Therefore butterfly singularity appears.

For the case(2), only when $n \le 3$, the lines of Eqs.(\ref{critII}) and 
(\ref{cuspIm}) are tangential to each other 
in the region Eq.(\ref{region}) and the value of $|Q_n|$ is

\begin{eqnarray}
	|\Qn|=|\Qn^{(2)}| &\equiv& 
	{\sqrt{n^2+9n+16+n\sqrt{n^2+10n+17}} \over 2n(n+1)(n+2)}
\nonumber \\
\label{qII}  
 & \qquad \times &
     \Bigl\{ {n(n+1+\sqrt{n^2+10n+17}) \over 4(n+2)} \Bigr\}^{n/2}~.
\end{eqnarray}

And then cusps merge to fold, $i.e.$, swallowtail singularity appears.

We surmmarize the result:
\begin{itemize}
\item[(I)] for $n=1$, $P_3=(\infty~,~\infty)$
\begin{itemize}
\item[(i)] $|\Qn|<|\Qn^{(0)}|$ \\
There are three solutions on $X=Y^2$ line and two solutions 
for $X>Y^2$, so $3+2 \times 2=7$ cusps on lens plane.
\item[(ii)] $|\Qn^{(0)}|<|\Qn|<|\Qn^{(2)}|$ \\
After beak-to-beak singularity appears, 
there is a solution on $X=Y^2$ line and two solutions 
for $X>Y^2$, so $1+2 \times 2=5$ cusps on lens plane.
\item[(iii)] $|\Qn^{(2)}|<|\Qn|$ \\
After swallowtail singularity appears, 
there is a solution on $X=Y^2$ line and no solution 
for $X>Y^2$, so one cusp on lens plane.
\end{itemize}
\item[(II)] for $n=2, 3$
\begin{itemize}
\item[(i)] $|\Qn|<|\Qn^{(0)}|$ \\
There are three solutions on $X=Y^2$ line and one solution
for $X>Y^2$, so $3n+2n=5n$ cusps on lens plane.
\item[(ii)] $|\Qn^{(0)}|<|\Qn|<|\Qn^{(1)}|$ \\
After beak-to-beak singularity appears, 
there is a solution on $X=Y^2$ line and one solution 
for $X>Y^2$, so $n+2n=3n$ cusps on lens plane.
\item[(iii)] $|\Qn^{(1)}|<|\Qn|<|\Qn^{(2)}|$ \\
After butterfly singularity appears, 
there is a solution on $X=Y^2$ line and two solutions 
for $X>Y^2$, so $n+2 \times 2n=5n$ cusps on lens plane.
\item[(iv)] $|\Qn^{(2)}|<|\Qn|$ \\
After swallowtail singularity appears, 
there is a solution on $X=Y^2$ line and no solution 
for $X>Y^2$, so $n$ cusps on lens plane.
\end{itemize}
\item[(III)] for $n \ge 4$
\begin{itemize}
\item[(i)] $|\Qn|<|\Qn^{(0)}|$ \\
There are three solutions on $X=Y^2$ line and one solution
for $X>Y^2$, so $3n+2n=5n$ cusps on lens plane.
\item[(ii)] $|\Qn^{(0)}|<|\Qn|<|\Qn^{(1)}|$ \\
After beak-to-beak singularity appears, 
there is a solution on $X=Y^2$ line and one solution 
for $X>Y^2$, so $n+2n=3n$ cusps on lens plane.
\item[(iii)] $|\Qn^{(1)}|<|\Qn|$ \\
After butterfly singularity appears, 
there is a solution on $X=Y^2$ line and no solution 
for $X>Y^2$, so $n$ cusps on lens plane.
\end{itemize}
\end{itemize}


\section{Two Point-Mass Model}

Critical curve produced by the lens galaxy of PG1115+080 makes us imagine
that the lens galaxy may be approximated by the two point-mass lens
[Fig.3].

\begin{center}
  ------------  \\
     Fig.3    \\
  ------------
\end{center}

\par
This model is interesting as its own right from the following reason.
In the previous section we studied the complex lens object by decomposing
it into $2^n$-pole contribution.
Two point-mass model is very simple one but is composed of composite
multipoles.
\par
So in this section we analyse this model in the analogous manner as in 
the previous section.
It is found that this model is qualitatively different from the MPE model.
\par
This model has been discussed by Erdl and Schneider \cite{Erdl}.
A straightforward application of catastrophe theory to this model still
enforces numerical calculations.
However we can derive singularities in simple analytical ways
by pursuing the critical
conditions of $f(r, \rho)$ defined in Eq.(\ref{fnoexp}).
\par
The surface mass density of two point-mass lens is given by

\begin{equation}
\Sigma(\vec x) = M[\mu_A \delta^2(\vec x - \vec a) 
                   + \mu_B \delta^2(\vec x + \vec a)]~~,
\end{equation}

here $M$ is total mass and $\mu_A + \mu_B = 1$.
$\vec a$ and $-\vec a$ are locations of two point masses in the lens surface.
\par
Fermat potential $\tilde \phi$ is given by

\begin{equation}
\tilde \phi = {1 \over 2}|z-z_s|^2
               - {m \over 2}(\mu_A \ln |z-\chi|^2 + \mu_B \ln |z+\chi|^2)~~,
\label{pofermat}
\end{equation}

in complex representation with $m \equiv {M \over \pi \Sigma_{cr}}$
and $\chi \equiv a_1 + ia_2$.
\par
We can rescale the potential into

\begin{equation}
\phi \equiv {\tilde \phi \over m}
= {1 \over 2}|u-v|^2 - {1 \over 2}[\mu_A \ln |u-\xi|^2 
                                    + \mu_B \ln |u+\xi|^2]
\end{equation}

up to irrelevant constant.
Here $u \equiv {z \over \sqrt{m}}$,
     $v \equiv {z_s \over \sqrt{m}}$~
and~ $\xi \equiv {\chi \over \sqrt{m}}$.                                    
\par
Lens equation ${\partial \phi \over \partial u}=0$ gives

\begin{equation}
\bar v = \bar u - ({\mu_A \over u-\xi} + {\mu_B \over u+\xi})~~.
\end{equation}


Critical curve which is given by 
$D={1 \over 4}-|\phi_{uu}|^2=0$
in general system of coordinates takes the following form,

\begin{equation}
|{\mu_A \over (u-\xi)^2}+{\mu_B \over (u+\xi)^2}|=1~.
\label{2pocrit}
\end{equation}

Introducing the new complex variable 
$U \equiv {u+\xi \over \xi}$,
$\xi$ in Eq.(\ref{2pocrit}) is factorized out as

\begin{equation}
|\xi|^2={|U^2-4\mu_B U+4\mu_B| \over |U (U -2)|^2}~.
\label{factorized}
\end{equation}

Using the polar coordinates 
$U = re^{i\theta}$,
critical curve Eq.(\ref{factorized}) is rewritten as

\begin{equation}
|\xi|^4=f(r,\cos \theta)~,
\label{fnodef}
\end{equation}

where

\begin{equation}
f(r,\rho) \equiv
{16\mu_B r^2 \rho^2-8\mu_B r(r^2+4\mu_B)\rho +r^4+8\mu_B (2\mu_B-1)r^2
 +16\mu_B^2 \over r^4(r^2-4r\rho +4)^2}
\label{fnoexp}
\end{equation}

with $\rho \equiv \cos \theta$.
\par
Beak-to-beak condition
${\partial D \over \partial U}=0$
becomes

\begin{equation}
{\partial f \over \partial U}=0
\label{bbcondi}
\end{equation}

from Eq.(\ref{fnodef}).
Eq.(\ref{bbcondi}) is equivalent to

\begin{equation}
\hbox{Re}({\partial f \over \partial U})
 =\rho {\partial f \over \partial r}
   +{1-\rho^2 \over r}{\partial f \over \partial \rho}=0
\label{bbreal}
\end{equation}

and

\begin{equation}
\hbox{Im}({\partial f \over \partial U})
 =\sin \theta ({\partial f \over \partial r}
   -{\rho \over r}{\partial f \over \partial \rho})=0~.
\label{bbimage}
\end{equation}

and, therefore, to

\begin{equation}
\hbox{case i)}~~~~~~~
{\partial f \over \partial r}={\partial f \over \partial \rho}=0
\label{case1}
\end{equation}

for $\sin \theta \ne 0$ or to

\begin{equation}
\hbox{case ii)}~~~~
\sin \theta ={\partial f \over \partial r}=0~.
\label{case2}
\end{equation}

Beak-to-beak appears at the points where four real parameters
$r$, $\rho$, $\mu_B$ and $|\xi|$
satisfy three equations (\ref{fnodef}) and (\ref{case1}) or (\ref{case2}).
\par
Thus metamorphoses of the caustics are appeared in the critical
behaviours of $f(r,\rho)$.
\par
Firstly we consider case i).

\begin{equation}
{\partial f \over \partial \rho}=
{8(1-\mu_B)(16\mu_B \rho -12\mu_B r+r^3) \over
 r^2 (r^2-4r\rho +4)^3}~.
\label{rhobibun}
\end{equation}

From the definitions,
$1-\mu_B >0$, $r^2-4r\rho +4>0$
($|\rho|<1$ for case i))
and $f$ takes the minimum value $F_m (r)$ at

\begin{equation}
\rho = {3 \over 4}r-{r^3 \over 16\mu_B} \equiv m(r)
\label{rhomin1}
\end{equation}

for fixed $r$.

\begin{equation}
F_m (r) \equiv f(r,m(r)) =
{\mu_B (r^2-4\mu_B)^2 \over r^4(r^4-8\mu_B r^2+16\mu_B)}~~(>0)~.
\label{Fmin1}
\end{equation}

We proceed to study the critical behaviour of $F_m (r)$.

\begin{equation}
{\partial F_m (r) \over \partial r}=
-{4\mu_B (r^2-4\mu_B) \over r^5(r^4-8\mu_B r^2+16\mu_B^2)}~g(r^2)~,
\label{gnodef}
\end{equation}

where

\begin{equation}
g(r^2) \equiv r^6-12 \mu_B r^4+48\mu_B^2 r^2-64\mu_B^2~.
\label{gnoexp}
\end{equation}

$g(r^2)$ is a monotonically increasing function of $r^2$ 
and the root of $g(r^2)=0$ is given by $r_1^2$,

\begin{equation}
r_1^2 =4\mu_B^{2/3}(\mu_A^{1/3}+\mu_B^{1/3}) \equiv 4\mu_B^{2/3} \delta~.
\label{gamma}
\end{equation}

The behaviour of $F_m (r)$ is depicted in Fig.4.

\begin{center}
  -------------  \\
     Fig.4    \\
  -------------
\end{center}

Nextly we consider case ii) where $\rho = \pm 1$.
We define $f_{\pm} \equiv f(r, \pm 1)$.
\par
Then

$$
f_{-}={(r^2+4\mu_B r+4\mu_B)^2 \over r^4(r+2)^4}
$$

is a monotonically decreasing function of $r$ and has no critical point.
On the other hand,

\begin{equation}
f_{+}={(r^2-4\mu_B r+4\mu_B)^2 \over r^4(r-2)^4}
\label{fplus}
\end{equation}

and

\begin{equation}
{df_{+} \over dr}=-{4 \over r^5(r-2)^5} (r^2-4\mu_B r+4\mu_B)~h(r)~,
\label{fplusnobibun}
\end{equation}

where

\begin{equation}
h(r) \equiv r^3-6\mu_B r^2+12\mu_B r-8\mu_B~.
\label{hnoexp}
\end{equation}

$h(r)$ is a monotonically increasing function of $r$ and has only one
real solution $r_2$ to $h(r)=0$,

\begin{equation}
r_2=2\mu_B^{1/3}(\mu_A^{2/3}+\mu_B^{2/3}-\mu_A^{1/3}\mu_B^{1/3})
   ={2\mu_B^{1/3} \over \delta}~.
\label{r2noexp}
\end{equation}

The behaviour of $f_{+}$ is depicted in Fig.5.

\begin{center}
  -------------  \\
     Fig.5    \\
  -------------
\end{center}

The above discussions in case i) and case ii) together with the facts that

\begin{eqnarray}
f(0, \rho) &=& \infty~~,~~f(\infty, \rho)=0~~\hbox{and} \\
f(2, 1) &=& \infty \nonumber
\label{facts}
\end{eqnarray}

are summarized as the contour map of $f(r, \rho)$ [Fig.6].

\begin{center}
  -------------  \\
     Fig.6    \\
  -------------
\end{center}

It is found that the critical curve changes its topology when we cross
the contour $L_1~(L_2)$ passing through $C_1~(C_2)$.
We denote the $\xi$ values corresponding to the contour $L_1~(L_2)$
as $\xi_1~(\xi_2)$.
Then

\begin{equation}
|\xi_1|^4=f|_{C_1}={1 \over 16\delta^3}
\label{xi1}
\end{equation}

and

\begin{equation}
|\xi_2|^4=f|_{C_2}={\delta^6  \over 16}~,
\label{xi2}
\end{equation}

which agree with the results of Erdl and Schneider \cite{Erdl}.

\section{The Application to the Multiple Quasar PG1115+080}

In this section, we apply our MPE model to the 
lensing galaxy of multiple quasar PG1115+080.
We use deflection potential up to the $2^3$-pole expansion here.
\par
About this lensed quasar,
we have got the observed data of positions and relative amplifications
of the images \cite{Chris}.
(We have adopted the new data for only $z_l$ from Angonim-Willaime et al.
 \cite{Angonim})
The positions of the images are as follows;

\begin{eqnarray}
 \theta(A_1) &=& (-0^{\prime \prime}.94, -0^{\prime \prime}.73) \nonumber \\ 
 \theta(A_2) &=& (-1^{\prime \prime}.11, -0^{\prime \prime}.27) \nonumber \\
 \theta(B) &=& (0^{\prime \prime}.72, -0^{\prime \prime}.60) \\
 \theta(C) &=& (0^{\prime \prime}.33, 1^{\prime \prime}.35) \nonumber 
\end{eqnarray}

And the relative amplifications of the images are as follows;

\begin{eqnarray}
 |\mu(A_1) / \mu(C)| &=& 3.22 \nonumber \\
 |\mu(A_2) / \mu(C)| &=& 2.49  \\
 |\mu(B) / \mu(C)| &=& 0.64  \nonumber 
\end{eqnarray}

Lens equation up to the $2^3$-pole expansion is given by

\begin{equation}
 v^*-u^*+{1 \over u}+{D \over u^2}+{2Q \over u^3}+{3T \over u^4}=0~.
\label{fiteq} 
\end{equation}

Here $v$ is source position, $u$ is image position, and $D$, $Q$, $T$
are dipole, quadrupole and $2^3$-pole, respectively.
These are all dimensionless complex numbered quantities normalized by 
suitable power of mass
(See Eq.(\ref{renormalize})).
Image amplification is given by 

\begin{equation}
\mu=|1-|{m \over z^2}+{2d \over z^3}+{6q \over z^4}
                   +{12t \over z^5}|^2|^{-1}~.
\end{equation}

We obtain values of the parameters included in Eq.(\ref{fiteq}) 
which reproduce the observed data well.
These parameter values are as follows;

\begin{eqnarray}
 \hbox{source position} &=& 
        (0^{\prime \prime}.02785, 0^{\prime \prime}.01507) \nonumber \\
 m &=& 3.496 \times 10^{-11}~~~[\hbox{rad}^2] \nonumber \\
 |D| &=& 0.1824~,~~~~~~~\theta_D=88^\circ .81  \\
 |Q| &=& 0.08724~,~~~~~~~\theta_Q=130^\circ .6  \nonumber \\
 |T| &=& 3.407 \times 10^{-3}~,~~~~~~~\theta_T=115^\circ .5 \nonumber
\label{para} 
\end{eqnarray}

If we assume Einstein-de Sitter universe, this mass would be
$M = \pi \Sigma_{cr} D_l^2 m  
   = 1.483 \pm 0.024~h^{-1}~\times~10^{11}~M_{\odot}$.
Here redshifts of the lensing galaxy and the source quasar
are $z_l = 0.294 \pm 0.005$ and  $z_s = 1.722$, respectively. 
Hubble constant is $H_0 = 100~h~$km/sec/Mpc.
\par
Calculated image positions are

\begin{eqnarray}
 \theta(A_1) &=& (-0^{\prime \prime}.9383, -0^{\prime \prime}.7317) 
                                                             \nonumber \\ 
 \theta(A_2) &=& (-1^{\prime \prime}.110, -0^{\prime \prime}.2748) 
                                                             \nonumber \\
 \theta(B) &=& (0^{\prime \prime}.7205, -0^{\prime \prime}.6004) \\
 \theta(C) &=& (0^{\prime \prime}.3294, 1^{\prime \prime}.350) \nonumber \\
 \theta(D) &=& (0^{\prime \prime}.5915, -0^{\prime \prime}.4450) \nonumber \\
 \theta(E) &=& (-0^{\prime \prime}.4494, 0^{\prime \prime}.08641)~. \nonumber
\end{eqnarray}

And calculated relative amplifications are

\begin{eqnarray}
 |\mu(A_1) / \mu(C)| &=& 3.113 \nonumber \\
 |\mu(A_2) / \mu(C)| &=& 2.892 \nonumber \\
 |\mu(B) / \mu(C)| &=& 0.7467 \\
 |\mu(D) / \mu(C)| &=& 0.1780 \nonumber \\ 
 |\mu(E) / \mu(C)| &=& 0.005871~. \nonumber 
\end{eqnarray}

Then, we should check the validity of approximation in multipole expansion
up to $2^3$-pole.
It must hold the following conditions.

\begin{equation}
 1 \gg {|D| \over |u|} \gg {|Q| \over |u|^2} \gg {|T| \over |u|^3}
\end{equation}

And the best fit parameters (\ref{para}) give following values.

\begin{equation}
{|D| \over |u|} \approx 0.2609~,~~{|Q| \over |u|^2} \approx 0.2103~,~~
   {|T| \over |u|^3} \approx 1.621 \times 10^{-2}
\end{equation}

This result roughly holds the condition 
${|Q| \over |u|^2} \gg {|T| \over |u|^3}$.
So this MPE model up to $2^3$-pole has been proved to be valid up to
$2^3$-pole moment.
\par
We performed parameter fitting of this lensed quasar
using MPE model up to quadrupole in the previous article
\cite{prev}.
The values of source position, dipole and quadrupole in Eq.(\ref{para})
are similar to those values in previous fitting.
So they are consistent.


\section{The Relations with Other Models}

We have shown in the previous sections that deflection potential up to
the quadrupole expansion reproduces multiple images well.
Its critical curve takes the similar form produced by 
two point-mass lens \cite{Erdl}.
\par
So it may be meaningful to check that two point-mass lens model on the single
lens plane well resembles with the lens of PG1115+080.
\par
Also it is very helpful for the study of lensing object that we analyze
the same lensing system by qualitatively different models.
As a typical example, we consider softened power-law spheroid models.

\subsection{Two point-mass lens}

Fermat potential given by Eq. (\ref{pofermat}) is expanded by multipoles as

\begin{equation}
\tilde \phi = {1 \over 2}|\vec x - \vec y|^2         
         - m \ln |\vec x| - {d_i x^i \over |\vec x|^2}
         - {q_{ij} x^i x^j \over |\vec x|^4} - ...~~.
\label{mulpsi}         
\end{equation}

Here

\begin{eqnarray}
\label{mulpara}
m &\equiv& {M \over \pi \Sigma_{cr}}  \nonumber \\
\label{muldef}
d_i &\equiv& m \sum_k \mu_k d_{ki}~~~~~~~~(k = A, B)  \\
q_{ij} &\equiv& m \sum_k \mu_k (d_{ki} d_{kj}
                  - {|\vec d_k|^2 \over 2} \delta_{ij}) \nonumber  
\end{eqnarray}

and so on, which are the real representations of Eqs.(\ref{fermul})
- (\ref{polemul}).
\par
In this case we have six independent parameters $M~, \mu_A~, \vec d_A~,
\vec d_B$ besides source positions $\vec y$.
Here we assume that MPE model up to quadrupole moments is
good approximation as we have shown 
and that two point-mass lens can be considered
to be their prototype.
Then we must be able to fit the parameters $M~, \mu_A~, \vec d_A~,
\vec d_B$ from
Eq.(\ref{mulpara}) so as to give the same $m~, d_i~, q_{ij}$ as the
MPE model.
This is always possible.
For Eq.(\ref{mulpara}) consists of five equations and we have six
parameters to be adjusted.
\par
Whether two point-mass model is good approximation or not, therefore,
depends on the fact that the higher multipole terms subsequent to
quadrupole in Eq.(\ref{mulpsi}) is small or not.
Here higher multipole terms is fixed by $M~, \mu_A~, \vec d_A~, \vec d_B$
determined by Eq.(\ref{muldef}).
\par
As is easily checked it is impossible.
Therefore two point-mass lens model cannot be a good approximation
of PG1115+080.


\subsection{Softened power-law spheroids (SPLS)}

This model is also called elliptical lens model whose mass density is 
given by

\begin{equation}
\rho(a) = \rho_0 (1+{a^2 \over r_c^2})^{-{k \over 2}}
\label{ellipdensity}
\end{equation}

with

\begin{equation}
x'^2 + y'^2 + {z'^2 \over 1-e^2} = a^2
\label{ellipradius}
\end{equation}

The principal axes of spheroid ($x',y',z'$) are related with the coordinate
of optical axis and lens surface ($x,y,z$) by

\begin{equation}
 \left(
   \begin{array}{c}
     x' \\ y' \\ z'
   \end{array} 
   \right)
 = \left(
   \begin{array}{ccc}
     1 & 0 & 0 \\
     0 & \cos \gamma & -\sin \gamma \\
     0 & \sin \gamma & \cos \gamma
   \end{array}
   \right)
   \left(
   \begin{array}{c}
     x \\ y \\ z
   \end{array}
   \right)~.
\end{equation}

That is, optical axis $z$ is tilted to $z'$ by $\gamma$.
\par
Here $r_c$ is the core radius and $\rho_0$ is the constant central density.
$e$ is the eccentricity.
For $k \le 3$ total mass is divergent, mass density has a cut-off

\begin{equation}
\rho(a) = 0~~~~~~~~~~~~~~~~~~~\hbox{for}~~~~{a \over r_c} > n~~.
\label{cut-off}
\end{equation}

This model has been applied to PG1115+080 by Narasimha et al. \cite{Narasimha}.
However they used the observational data different from ours,
therefore we cannot simply compare their results with ours.
Elliptical lens model is qualitatively different model from MPE model.
The former is transparent and the latter non-transparent.
If we are only satisfied by reproducing image positions,
both model may pass our requirements.
So we must try parameter fitting to more various and precise data.
\par
The detail of this model has been discussed in a separate form 
\cite{Katsukun}, so we just quote the result 
[Table 1, Table 2].

\begin{center}
  --------------  \\
     Table 1      \\
  --------------
\end{center}

\begin{center}
  --------------  \\
     Table 2      \\
  --------------
\end{center}

SPLS has a large fitting zone for its transparency compared with MPE.
\par
As a whole this model gives rather non-compact lensing object 
unlike the compact
aspect of Narasimha et al. \cite{Narasimha}.
This seems to be incompatible with the observed lensing object
\cite{Kris}.
Observed image amplifications are better fitted by MPE
model than by elliptical lens model.



\section{Discussions}

MPE model is not so popular as the elliptical lens model
with or without external shear
since most of the gravitational lensings are thought to be caused by
the transparent global systems composed of many galaxies.
However, the recent observations support that the lens of PG1115+080 is
a compact object whose optical extent is within the observed innermost
image B $\sim 10 h^{-1}$ kpc \cite{Chris} \cite{Kris}.
MPE model is neither transparent nor nonsingular.
Therefore it is free from the general theorems about gravitational lensing
(See chapter 5.4 of Schneider et al. \cite{Schneider}).
Indeed, in Fig.3 we have 4 images of odd parity and 2 images of even parity.
There gives rise, of course, no inconsistency.
\par
We must wait further refinement of model check in order that our model
parameters are physically affirmative.
For that purpose, we need more precise observations such as
image shape, time delays and spectroscopy profile 
\cite{Micha}, etc.
In this connection it is in order to comment on time delay.
\par
Time delay is given by

\begin{equation}
t = {1 \over c} (1+z_l) {D_l D_s \over D_{ls}} \phi~~.
\end{equation}

For $\Omega_0 = 1$ case,

\begin{eqnarray}
D_{ij} &=& {2c \over H_0}
  {(1+z_j)\sqrt{1+z_i}-(1+z_i)\sqrt{1+z_j} \over (1+z_j)^2 (1+z_i)}~~. \\
&(& z_i < z_j~,~~ i, j = \hbox{source,~lens,~observer}~~) \nonumber 
\end{eqnarray}

For $\Omega_0 = 0$ case,

\begin{equation}
D_{ij} = {c \over 2H_0}
  {(1+z_j)^2 - (1+z_i)^2 \over (1+z_j)^2 (1+z_i)}~~.
\end{equation}

Time delays for each case are given in Table 3.

\begin{center}
  --------------  \\
     Table 3      \\
  --------------
\end{center}

For comparison we have also listed the respective time delay
based on the elliptical lens model [Table 2].
Estimations by MPE model are roughly coincident with the 
observation by Vanderriest et al.\cite{Vanderriest}.
On the contrary,
estimations by SPLS in Table 2 are too short,
which denies the possibility that its amplification discrepancy 
with observation might be remedied by the time variation of amplification 
of source QSO.
Naively we may conclude that the lensing object of PG1115+080 is compact
also from the theoretical side.
\par
Consequently MPE model deserves serious surveys.
In this connection
it should be remarked that the dipole term plays an important role
in our theory :
The presence of dipole moment
means that the origin of the coordinates
(which is the optical center)
is away from the center of
mass of the lensing galaxy.
There are the arguments which insist that this deviation may be
the cause of spiral arms.
Also MPE model gives better fit to the observational data
than elliptical lens model in many points \cite{Katsukun}.
This may supports the view point that the lensing galaxy is a spiral galaxy,
whose equipotential contour is different from spheroid.
There is the observational support of this viewpoint from the analysis
of spectroscopy \cite{Turner}\cite{Micha}.
From these facts, non zero dipole component may play any role in
the formation of spiral arms. 
\par
Though we applied it only to PG1115+080 in this article,
MPE model is very useful for the wide class of lensing phenomena.
It gives systematically the multipole moments of the lensing object,
irrespective to isolated lens or not.
This serves crucially to the construction of more concrete and more precise
structure of lensing object.



\section*{Acknowledgements}

We are grateful to Dr. Hiroshi Yoshida for useful comments.

\vfill\eject

\vfill\eject


\begin{center}
{\bf Figure Captions}
\end{center}

\begin{description}
\item[Fig.1] The region that satisfies the cusp condition.
             Solid line is $X=Y^2$, dotted line is Re(cusp)=0
             and dashed line is Im(cusp)=0.
             Three lines pass through the points $P_1$ and $P_2$.
             The cusp condition is satisfied in the region 
             where the dashed line is in the common region of
             the interiors of the solid and the dotted line.
             Between $P_1$ and $P_2$, the dashed line passes very
             close to but in the exterior of the dotted line.

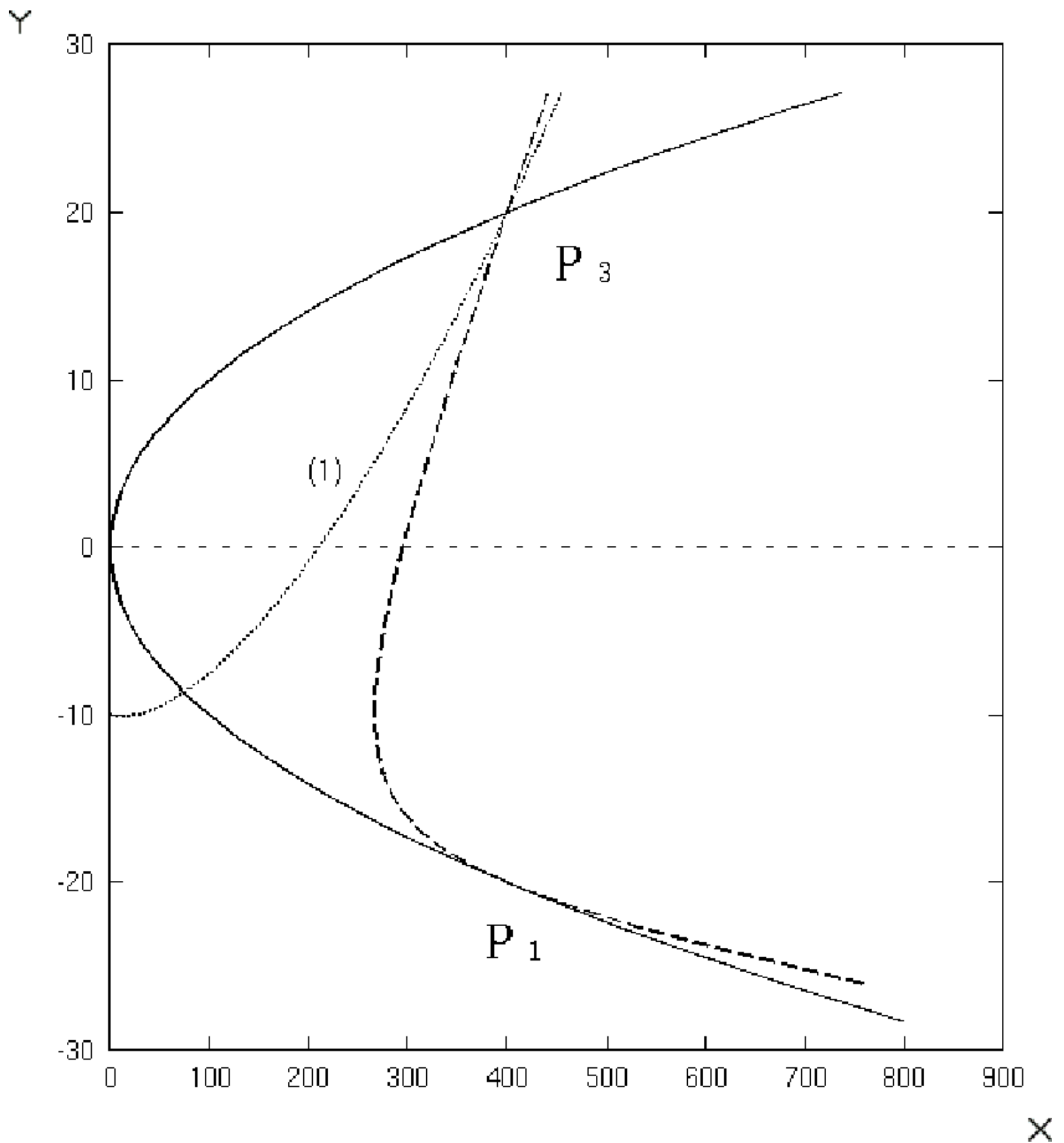
\item[Fig.2] The behaviours of the critical condition.
             Solid line is $X=Y^2$, dashed line is Im(cusp)=0 and
             dotted line is the critical condition.
             Fig.2(a) is the $n=2$ case.
             The dotted lines of (0), (1) and (2) correspond to
             the cases (0), (1) and (2) discussed in the text, respectively.
             Fig.2(b) is the $n=4$ case.

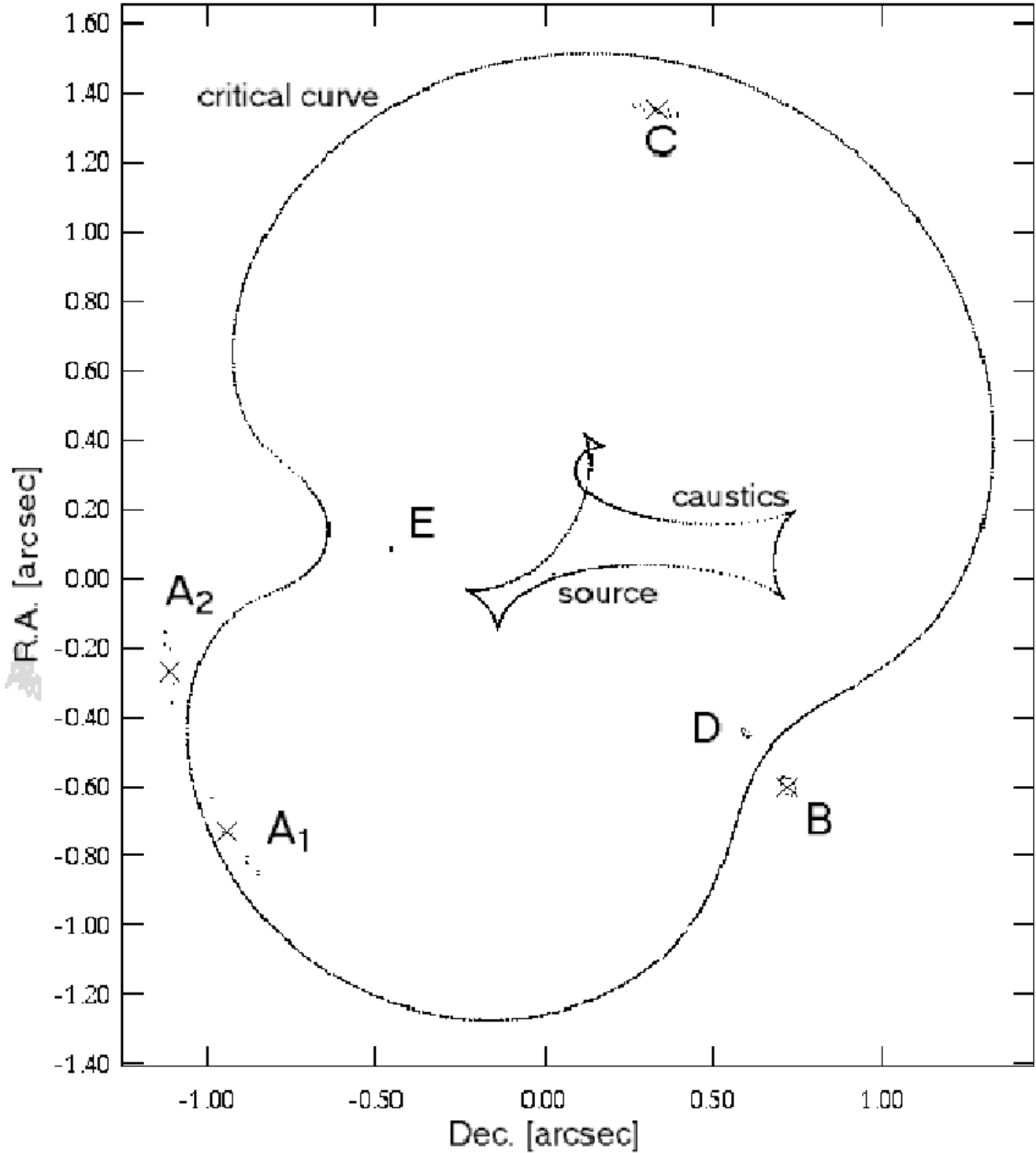
\item[Fig.3] Shapes of images, critical line and caustics
               for the multiple quasar PG1115+080 (Kakigi et al. 1995).
               Crosses (dotted regions) are the observed (calculated)
               image positions.
               It has been assumed that source QSO is spherical.

\item[Fig.4] The critical behaviour of $F_m(r)$.
               Critical points $C_0$ and $C_1$ in $(r, \rho)$ plane are
               $(r_0, \rho_0) \equiv (2\mu_B^{1/2}, \mu_B^{1/2})$ and
               $(r_1, \rho_1) \equiv (2\mu_B^{1/3} \delta^{1/2},
                     {\delta^{1/2} \over 2}(2\mu_B^{1/3}-\mu_A^{1/3}))$,
               respectively.
               We can set $\mu_B \geq \mu_A$ without spoiling
               generality and therefore $1 > \rho_1 > 0$.

\item[Fig.5] The critical behaviour of $f_{+}$.
               Critical point $C_2$ is given by
               $C_2=(r_2,\rho_2)=({2\mu_B^{1/3} \over \delta},1)$.

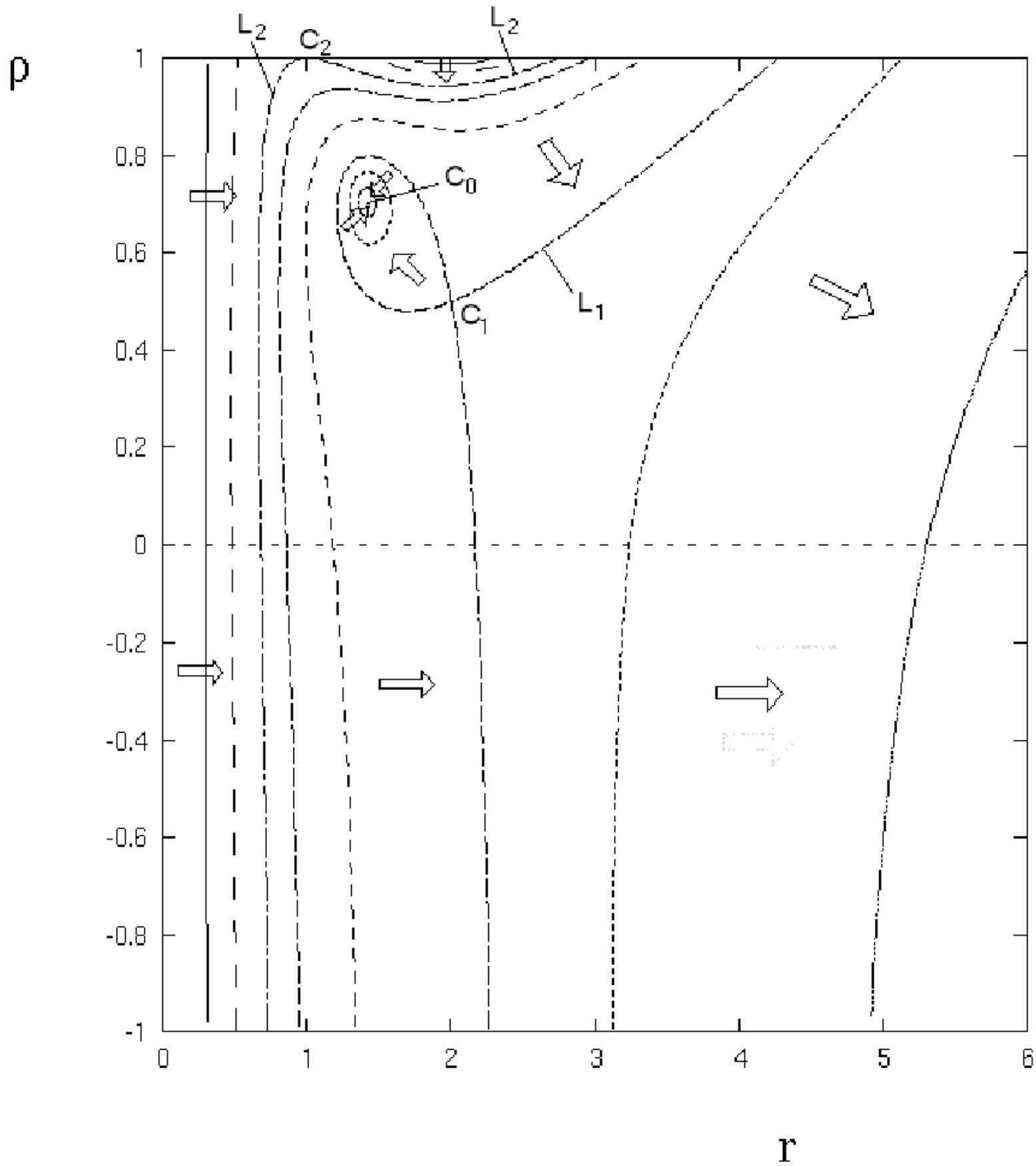
\item[Fig.6] The contour map of $f(r,\rho)$.
               Origin is the point of heavier point mass.
               $f(r,\rho)$ diminishes in the direction of $\Rightarrow$.
               $C_1$ and $C_2$ are the points where beak-to-beaks appear.

\item[Table 1] Fitting parameters of SPLS ;
               (a) k=3 case and (b) k=2 case.

\item[Table 2] Calculated image positions
                 (and time delays lagging behind C in $h^{-1}$ days)
                  of SPLS ;
               (a) k=3 case and (b) k=2 case.

\item[Table 3] Time delays lagging behind C in $h^{-1}$ days of MPE model.

\end{description}

\vfill\eject




\thispagestyle{empty}

\begin{center}
\begin{tabular}{|c|c|c|c|c|c|c|}
 \multicolumn{7}{l}{~~~~~~~~~~~~~~~~~~~~~~$C_0$~~~~~~~~~~~~$C_1$} \\ \hline
 $r$ & 0 &   & $2\mu_B^{1/2}$ &  & $r_1$ &   \\ \hline
 ${dF_m \over dr}$ &  & $-$ & 0 & $+$ & 0 & $-$  \\ \hline
 $F_m$ & $\infty$ & $\searrow$ & 0 & $\nearrow$ & 
 ${1 \over 16\delta^3}$ & $\searrow$ \\ \hline
\end{tabular}
\end{center}

\begin{center}
Fig.4
\end{center}

\begin{center}
\begin{tabular}{|c|c|c|c|c|c|c|}
 \multicolumn{7}{c}{$C_2$} \\ \hline
 $r$ & 0 &   & $r_2$ &  & 2 &   \\ \hline
 ${df_{+} \over dr}$ &  & $-$ & 0 &  &  & $-$  \\ \hline
 $f_{+}$ & $\infty$ & $\searrow$ & ${\delta^6 \over 16}$ &
 $\nearrow$ &  $\infty$ & $\searrow$ \\ \hline
\end{tabular}
\end{center}

\begin{center}
Fig.5
\end{center}

\vfill\eject



\thispagestyle{empty}

\begin{center}
\begin{tabular}{|c|c|c|c|c|c|}\hline
 $e \sin \gamma$ & $n$ & $r_c$ &
 $\kappa_0 \sqrt{1-e^2}$ &
 Total Mass & Source Position  \\ \hline\hline
 0.4 & 5 & 6.13~$h^{-1}$ kpc & 2.58$\times 10^{-23}~h^2 g / cm^3$ &
 1.48$\times 10^{12} M_\odot$ & 
 $(0^{\prime \prime}.024,~-0^{\prime \prime}.037)$  \\ \hline
 0.4 & 10 & 6.07~$h^{-1}$ kpc & 2.56$\times 10^{-23}~h^2 g / cm^3$ &
 2.13$\times 10^{12} M_\odot$ & 
 $(0^{\prime \prime}.024,~-0^{\prime \prime}.037)$  \\ \hline\hline
 0.5 & 5 & 4.78~$h^{-1}$ kpc & 3.48$\times 10^{-23}~h^2 g / cm^3$ &
 9.43$\times 10^{11} M_\odot$ &
 $(0^{\prime \prime}.038,~-0^{\prime \prime}.058)$  \\ \hline
 0.5 & 10 & 4.71~$h^{-1}$ kpc & 3.50$\times 10^{-23}~h^2 g / cm^3$ &
 1.37$\times 10^{12} M_\odot$ & 
 $(0^{\prime \prime}.039,~-0^{\prime \prime}.058)$  \\ \hline\hline
 0.6 & 5 & 3.84~$h^{-1}$ kpc & 4.66$\times 10^{-23}~h^2 g / cm^3$ &
 6.56$\times 10^{11} M_\odot$ &
 $(0^{\prime \prime}.057,~-0^{\prime \prime}.085)$  \\ \hline
 0.6 & 10 & 3.75~$h^{-1}$ kpc & 4.74$\times 10^{-23}~h^2 g / cm^3$ &
 9.30$\times 10^{11} M_\odot$ &
 $(0^{\prime \prime}.057,~-0^{\prime \prime}.084)$  \\ \hline
\end{tabular}
\end{center}

\begin{center}
Table 1~(a)
\end{center}

\begin{center}
\begin{tabular}{|c|c|c|c|c|c|}\hline
 $e \sin \gamma$ & $n$ & $r_c$ &
 $\kappa_0 \sqrt{1-e^2}$ &
 Total Mass & Source Position  \\ \hline\hline
 0.3 & 5 & 6.00~$h^{-1}$ kpc & 1.79$\times 10^{-23}~h^2 g / cm^3$ &
 2.62$\times 10^{12} M_\odot$ & 
 $(0^{\prime \prime}.014,~-0^{\prime \prime}.020)$  \\ \hline
 0.3 & 10 & 5.46~$h^{-1}$ kpc & 1.87$\times 10^{-23}~h^2 g / cm^3$ &
 4.83$\times 10^{12} M_\odot$ & 
 $(0^{\prime \prime}.014,~-0^{\prime \prime}.021)$  \\ \hline\hline
 0.4 & 5 & 4.23~$h^{-1}$ kpc & 2.71$\times 10^{-23}~h^2 g / cm^3$ &
 1.39$\times 10^{12} M_\odot$ &
 $(0^{\prime \prime}.025,~-0^{\prime \prime}.036)$  \\ \hline
 0.4 & 10 & 3.97~$h^{-1}$ kpc & 2.71$\times 10^{-23}~h^2 g / cm^3$ &
 2.70$\times 10^{12} M_\odot$ & 
 $(0^{\prime \prime}.025,~-0^{\prime \prime}.035)$  \\ \hline\hline
 0.5 & 5 & 3.13~$h^{-1}$ kpc & 3.93$\times 10^{-23}~h^2 g / cm^3$ &
 8.15$\times 10^{11} M_\odot$ &
 $(0^{\prime \prime}.039,~-0^{\prime \prime}.055)$  \\ \hline
 0.5 & 10 & 2.84~$h^{-1}$ kpc & 4.13$\times 10^{-23}~h^2 g / cm^3$ &
 1.51$\times 10^{12} M_\odot$ &
 $(0^{\prime \prime}.040,~-0^{\prime \prime}.054)$  \\ \hline
\end{tabular}
\end{center}

\begin{center}
Table 1~(b)
\end{center}

\vfill\eject
\thispagestyle{empty}

\begin{center}
\begin{tabular}{|c|c|c|c|c|c|}\hline
 $e \sin \gamma$ & $n$ & $A_1$ & $A_2$ & $B$ & $C$  \\ \hline\hline
 0.4 & 5 & $(0^{\prime \prime}.51,~1^{\prime \prime}.08)$~(4.0) &
 $(0^{\prime \prime}.95,~0^{\prime \prime}.63)$~(4.1) &
 $(-0^{\prime \prime}.88,~0^{\prime \prime}.31)$~(6.0) &
 $(0^{\prime \prime}.23,~-1^{\prime \prime}.37)$  \\ \hline
 0.4 & 10 & $(0^{\prime \prime}.51,~1^{\prime \prime}.08)$~(3.9) &
 $(0^{\prime \prime}.96,~0^{\prime \prime}.62)$~(4.0) &
 $(-0^{\prime \prime}.89,~0^{\prime \prime}.31)$~(5.9) &
 $(0^{\prime \prime}.24,~-1^{\prime \prime}.37)$  \\ \hline\hline
 0.5 & 5 & $(0^{\prime \prime}.53,~1^{\prime \prime}.06)$~(6.1) &
 $(0^{\prime \prime}.97,~0^{\prime \prime}.61)$~(6.3) &
 $(-0^{\prime \prime}.89,~0^{\prime \prime}.30)$~(9.4) &
 $(0^{\prime \prime}.24,~-1^{\prime \prime}.37)$  \\ \hline
 0.5 & 10 & $(0^{\prime \prime}.53,~1^{\prime \prime}.06)$~(6.2) &
 $(0^{\prime \prime}.97,~0^{\prime \prime}.61)$~(6.3) &
 $(-0^{\prime \prime}.89,~0^{\prime \prime}.30)$~(9.6) &
 $(0^{\prime \prime}.25,~-1^{\prime \prime}.37)$  \\ \hline\hline
 0.6 & 5 & $(0^{\prime \prime}.53,~1^{\prime \prime}.05)$~(9.0) &
 $(1^{\prime \prime}.00,~0^{\prime \prime}.57)$~(9.2) &
 $(-0^{\prime \prime}.89,~0^{\prime \prime}.28)$~(13.9) &
 $(0^{\prime \prime}.26,~-1^{\prime \prime}.37)$  \\ \hline
 0.6 & 10 & $(0^{\prime \prime}.53,~1^{\prime \prime}.05)$~(8.8) &
 $(1^{\prime \prime}.00,~0^{\prime \prime}.57)$~(9.1) &
 $(-0^{\prime \prime}.89,~0^{\prime \prime}.28)$~(13.8) &
 $(0^{\prime \prime}.26,~-1^{\prime \prime}.37)$  \\ \hline
\end{tabular}
\end{center}

\begin{center}
Table 2~(a)
\end{center}

\begin{center}
\begin{tabular}{|c|c|c|c|c|c|}\hline
 $e \sin \gamma$ & $n$ & $A_1$ & $A_2$ & $B$ & $C$  \\ \hline\hline
 0.3 & 5 & $(0^{\prime \prime}.52,~1^{\prime \prime}.08)$~(2.1) &
 $(0^{\prime \prime}.94,~0^{\prime \prime}.64)$~(2.1) &
 $(-0^{\prime \prime}.88,~0^{\prime \prime}.32)$~(3.2) &
 $(0^{\prime \prime}.23,~-1^{\prime \prime}.37)$  \\ \hline
 0.3 & 10 & $(0^{\prime \prime}.56,~1^{\prime \prime}.04)$~(2.2) &
 $(0^{\prime \prime}.92,~0^{\prime \prime}.67)$~(2.3) &
 $(-0^{\prime \prime}.90,~0^{\prime \prime}.31)$~(3.4) &
 $(0^{\prime \prime}.23,~-1^{\prime \prime}.35)$  \\ \hline\hline
 0.4 & 5 & $(0^{\prime \prime}.54,~1^{\prime \prime}.06)$~(3.9) &
 $(0^{\prime \prime}.96,~0^{\prime \prime}.62)$~(4.0) &
 $(-0^{\prime \prime}.88,~0^{\prime \prime}.31)$~(6.1) &
 $(0^{\prime \prime}.26,~-1^{\prime \prime}.36)$  \\ \hline
 0.4 & 10 & $(0^{\prime \prime}.52,~1^{\prime \prime}.07)$~(3.8) &
 $(0^{\prime \prime}.92,~0^{\prime \prime}.61)$~(4.0) &
 $(-0^{\prime \prime}.89,~0^{\prime \prime}.30)$~(6.1) &
 $(0^{\prime \prime}.26,~-1^{\prime \prime}.36)$  \\ \hline\hline
 0.5 & 5 & $(0^{\prime \prime}.54,~1^{\prime \prime}.05)$~(5.8) &
 $(0^{\prime \prime}.99,~0^{\prime \prime}.58)$~(6.0) &
 $(-0^{\prime \prime}.89,~0^{\prime \prime}.29)$~(9.3) &
 $(0^{\prime \prime}.28,~-1^{\prime \prime}.36)$  \\ \hline
 0.5 & 10 & $(0^{\prime \prime}.54,~1^{\prime \prime}.05)$~(5.8) &
 $(0^{\prime \prime}.99,~0^{\prime \prime}.57)$~(6.0) &
 $(-0^{\prime \prime}.89,~0^{\prime \prime}.29)$~(9.4) &
 $(0^{\prime \prime}.28,~-1^{\prime \prime}.36)$  \\ \hline
\end{tabular}
\end{center}

\begin{center}
Table 2~(b)
\end{center}

\vfill\eject
\thispagestyle{empty}

\begin{center}
\begin{tabular}{|l|c|c|c|c|c|}\hline
 $\Delta$~($h^{-1}$ days) &
 $C-A_1$ & $C-A_2$ & $C-B$ & $C-D$ &  $C-E$ \\ \hline\hline
 $\Omega_0=1$ & 
 $9.757 \pm 0.200$ & $11.448 \pm 0.234$ & $19.407 \pm 0.396$ &
 $19.225 \pm 0.392$ & $3.076 \pm 0.063$ \\ \hline
 $\Omega_0=0$ &
 $10.540^{+0.227}_{-0.225}$ & $12.367^{+0.266}_{-0.264}$ & 
 $20.964^{+0.453}_{-0.447}$ &
 $20.767^{+0.448}_{-0.443}$ & $3.323^{+0.072}_{-0.071}$ \\ \hline
\end{tabular}
\end{center}

\begin{center}
Table 3
\end{center}

\vfill\eject

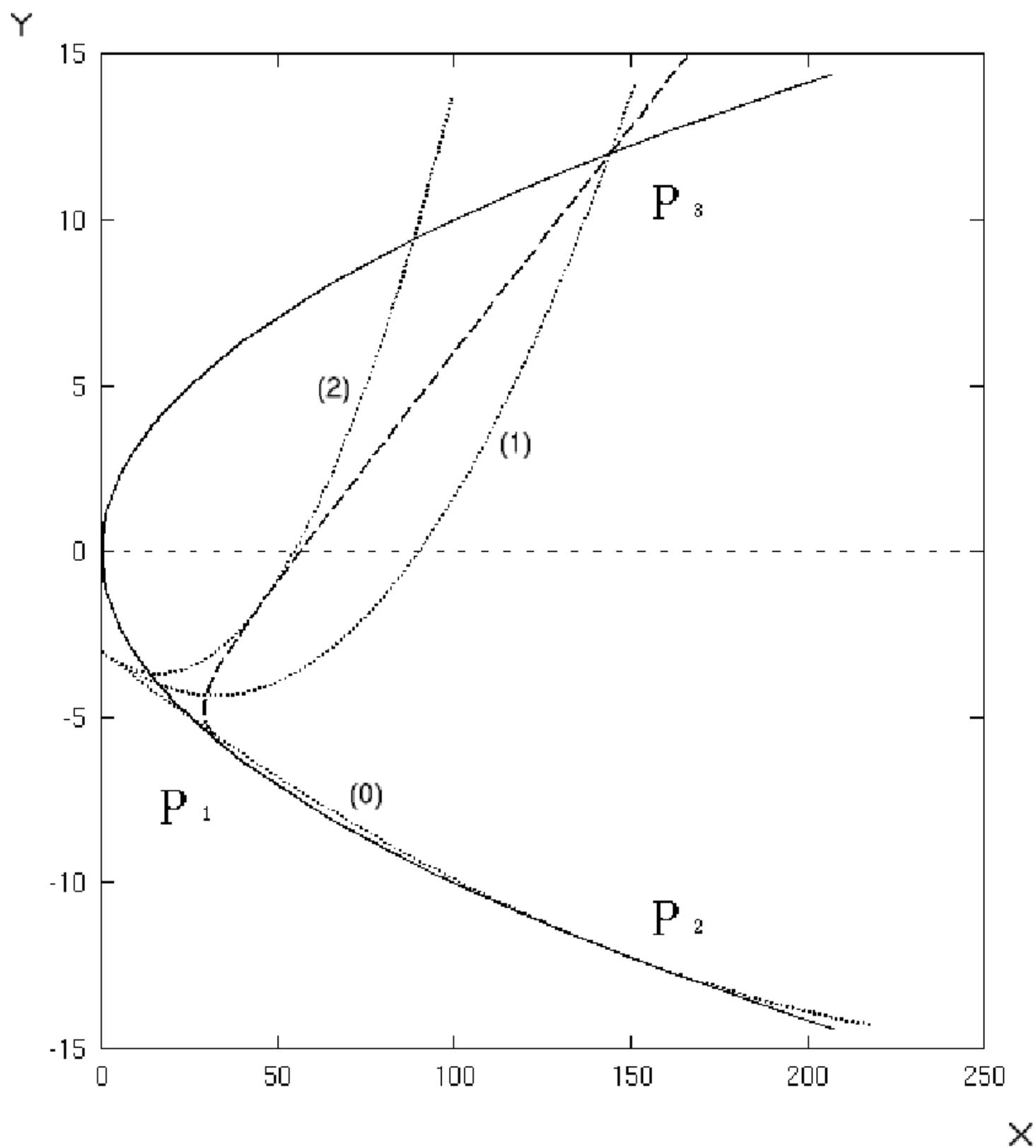
\end{document}